\documentclass[twocolumn,floatfix, nofootinbib,prd,preprintnumbers,showpacs,showkeys]{revtex4-1}

\usepackage[mathscr]{euscript}
\usepackage{amsmath}
\usepackage{graphicx}
\usepackage{dcolumn}
\usepackage{bm}
\usepackage{epsfig}
\usepackage{amssymb,latexsym,mathrsfs}
\usepackage{graphicx}
\usepackage{color}
\usepackage{hyperref}
\usepackage{float}
\usepackage{diagbox}
\usepackage{mathtools}

\usepackage{tikz}
\usepackage[compat=1.1.0]{tikz-feynman}

\usepackage{amsmath}
\newcommand*\diff{\mathop{}\!\mathrm{d}}

\hypersetup{
    colorlinks=true,
    linkcolor=red,
    citecolor=blue,
}

\usepackage{amsmath}
\usepackage{amssymb}
\usepackage{subfigure}
\usepackage{hyperref}
\usepackage{url}
\usepackage{xcolor}
\usepackage{color}
\definecolor{amaranth}{rgb}{0.9, 0.17, 0.31}
\definecolor{purple(munsell)}{rgb}{0.62, 0.0, 0.77}
\definecolor{americanrose}{rgb}{1.0, 0.01, 0.24}
\definecolor{palatinateblue}{rgb}{0.15, 0.23, 0.89}
\definecolor{royalblue(web)}{rgb}{0.25, 0.41, 0.88}
\definecolor{hanpurple}{rgb}{0.32, 0.09, 0.98}
\definecolor{beaublue}{rgb}{0.74, 0.83, 0.9}
\definecolor{carminered}{rgb}{1.0, 0.0, 0.22}
\definecolor{brightpink}{rgb}{1.0, 0.0, 0.5}
\definecolor{vividviolet}{rgb}{0.62, 0.0, 1.0}
\hypersetup{ linktoc=all,
    colorlinks, linkcolor={palatinateblue},
    citecolor={brightpink}, urlcolor={amaranth}}

\newcommand{\be}{\begin{equation}}
\newcommand{\ee}{\end{equation}}
\newcommand{\bs}{\begin{split}} 
\newcommand{\bea}{\begin{eqnarray}}
\newcommand{\eea}{\end{eqnarray}}

\newcommand{\kap}{\kappa} 
\newcommand{\kp}{\kappa}

\newcommand{\bes}{\begin{subequations}}
\newcommand{\ees}{\end{subequations}}

\renewcommand{\d}[1]{\ensuremath{\operatorname{d}\!{#1}}}

\renewcommand{\d}[1]{\ensuremath{\operatorname{d}\!{#1}}}

\newcommand{\bo}{\raise-1mm\hbox{\Large$\Box$}}

\newcommand{\bd}{\boldsymbol}

\begin{document}

\title{Stopping to Reflect: 
Asymptotic Static Moving Mirrors as Quantum Analogs of Classical Radiation}

\author{Michael R.R. Good${}^{1,2,3}$}
\email{michael.good@nu.edu.kz}
\author{Eric V. Linder${}^{3,4}$}
\email{evlinder@lbl.gov}
\affiliation{${}^1$Physics Department, Nazarbayev University,
Astana, Kazakhstan\\
${}^2$ Leung Center for Cosmology and Particle Astrophysics,
National Taiwan University, Taipei, Taiwan\\
${}^3$Energetic Cosmos Lab, Nazarbayev University, 
Astana, Kazakhstan\\
${}^4$Berkeley Center for Cosmological Physics \& Berkeley Lab, University of California, Berkeley, CA, USA 
}

\begin{abstract} 
Radiation from an accelerating charge is a basic process 
that can serve as an intersection between classical and 
quantum physics. We present 
two exactly soluble electron trajectories that permit analysis of the radiation emitted, exploring its time evolution and spectrum by analogy with the moving mirror model of the dynamic Casimir effect.  
These classical solutions are finite energy, rectilinear (nonperiodic),  asymptotically zero velocity worldlines with corresponding quantum analog beta Bogolyubov coefficients. 
One of them has an interesting connection to uniform 
acceleration and Leonardo da Vinci's water pitcher experiment. 
\end{abstract} 

\date{\today} 

\maketitle


\section{Introduction} 

The mechanism of particle creation proposed by Hawking \cite{Hawking:1974sw}, whereby the gravitational field of a collapsing star in curved spacetime amplifies vacuum fluctuations into particle emission, bears striking resemblance to the radiation of particles from a perfect mirror in flat spacetime accelerated through the vacuum \cite{DeWitt:1975ys,Davies:1976hi,Davies:1977yv}.  
Particles of a massless quantum scalar field in $1+1$ dimensions \cite{Birrell:1982ix,Fabbri} are created due to the acceleration of the mirror, which is an ideal point and boundary condition on the field \cite{walker1985particle,Ford:1982ct,carlitz1987reflections,Chen:2015bcg,AnaBHEL:2022sri,Good:2021asq,Chen:2020sir}, essentially a dynamical Casimir effect \cite{moore1970quantum}. 
In this study, we demonstrate a functional duality and analog to an accelerated point charge 
in ordinary 3+1 space-time and its 
non-thermal radiation spectrum, revealing the particle creation correspondence.  

Accelerating point charge radiation has been a subject 
of interest in physics for over a century \cite{Larmor1897}, and it is of particular interest 
as a simple example of nonthermal radiation. 
Nonthermal radiation is ubiquitous in astrophysical 
phenomena, for example, and the particle number  
and angular spectral distribution may not be 
apparent. Furthermore, even 
evaporating black holes might emit non-thermal radiation, e.g.\ the recent \cite{Svidzinsky:2023yvk}. 
Therefore a concrete relation between accelerated 
particle nonthermal radiation and the moving mirror 
``slicing'' of the vacuum \cite{Lin:2021bpe,Reyes:2021npy,Lin:2022sbi,Akal:2020twv}, especially in light of 
the well-established correspondence between moving 
mirrors and black hole horizons, is of interest.

The discovery of a 
clear association (generalized to non-thermal emissions) between the radiation from an electron and from a moving mirror became apparent via radiation reaction derived by Ford and Vilenkin in 1982 \cite{Ford:1982ct}. In 1995, Nikishov and Ritus \cite{Nikishov:1995qs} established a formal link through particle count, which further strengthened this connection. Ritus \cite{Ritus:1999eu,Ritus:2002rq,Ritus:2003wu,Ritus:2022bph} later provided additional development on the Bogolyubov-current association. The relationship was next confirmed via Larmor power in Zhakenuly et al \cite{Zhakenuly:2021pfm}. One of the present authors has exploited the electron-mirror connection using explicit solutions; for instance, the connection between radiation power loss and kinetic power loss for an electron approaching the speed of light was demonstrated in \cite{Good:2022gvk}, and in \cite{Good:2023hsv} an electron was treated as a mirror for a trajectory that asymptotically approaches a constant velocity. 
This article focuses on the interesting results for 
the electron-mirror relation for 
trajectories that come to a complete stop, giving 
finite energy, finite particle creation, and 
unitary evolution.  

In Sec.\ \ref{sec:accrad}, we review some elements of acceleration radiation for relativistic moving point charges, including Larmor power, Feynman power, and 
their connection to total energy emitted. We present the spectra for two different motions of point charges and the quantum analogs that have desirable 
properties and analytic Bogolyubov coefficients 
in Sec.~\ref{sec:selfdual} and Sec.~\ref{sec:betaK}. 
In Sec.~\ref{sec:disc} we show the general correspondence between 
the classical bremsstrahlung and dynamical Casimir effect 
in energy, particle count, and spectral distribution. 
We summarize and discuss further areas for study in 
Sec.~\ref{sec:concl}.

\section{Acceleration Radiation Elements} \label{sec:accrad} 

In this section, we set up the various elements 
needed to compute the radiated power, energy, 
and spectral distribution of both an accelerating 
charge and from a moving mirror dynamical 
Casimir effect. 
Throughout we use natural units, 
$\hbar = c = \mu_0 = \epsilon_0 = 1$, thus $e^2 = 4\pi \alpha_{\rm fs}$ where $\alpha_{\rm fs}$ is the fine structure constant. However, for simplicity, when exclusively in the context of classical electrodynamics we switch units and employ unit charge $e=1$ ($\hbar = 1/4\pi\alpha_{\rm fs}$).

\subsection{Power and Force} 

In classical electrodynamics \cite{Jackson:490457}, the power radiated 
and the radiation reaction force,
\be P = \frac{\alpha^2}{6\pi}\ , \qquad F = \frac{\alpha'(\tau)}{6\pi}\ , \ee
are given by the relativistically covariant Larmor formula and the (magnitude of the) Lorentz-Abraham-Dirac (LAD) force.  Here $\alpha$ is the proper acceleration, and the prime is a derivative with respect to the argument, in this case  proper time  $\tau$.

\subsection{Energy Integrals} 

When the charged particle accelerates, energy is radiated, 
with the total energy found by integrating over coordinate time.  That is, for particle velocity $v(t)$ the integrals
\be 
E = \int_{-\infty}^{\infty} P \diff{t} = -\int_{-\infty}^{\infty} F\cdot v \diff{t}, \label{FL}
\ee
demonstrate that the Larmor power, $P= \alpha^2/6\pi$, and what we call the `Feynman power' \cite{Feynman:1996kb}, $F \cdot v$, associated with the self-force (radiation reaction force), 
directly tell an observer the total energy emitted by a point charge along its time-like worldline.  The total energy is finite as long as the proper acceleration is asymptotically zero; that is, the worldline must possess asymptotic inertia. We restrict ourselves to this case. 

The negative sign demonstrates that the total work against the LAD force represents the total energy loss. That is, the total energy loss from radiation resistance due to Feynman power must equal the total energy radiated by Larmor power.  We will demonstrate that the Larmor and Feynman powers themselves -- the integrands -- are not the same. Separately, it is a subtle matter that these powers are not applicable for  
asymptotically {\it non-inertial\/} rectilinear trajectories (which we do not consider here); see e.g.\  \cite{Good:2022gvk,Singal:2020kan}.  

A third expression for the total energy can be employed to establish a link to 
quantum physics and verify consistency. This spectral consistency integrates 
over spectral modes, 
\be 
E = \int_0^\infty \int_0^\infty p\, |\beta_{pq}|^2 \diff{p}\diff{q}\ , \label{energyfrombeta}
\ee
using the quantum analog moving mirror model, generalized to 3+1 dimensions using both sides of the 1+1 dimensional moving mirror, see e.g. \cite{Zhakenuly:2021pfm,Nikishov:1995qs}. 

The quantity $\beta_{pq}$ is the 
beta Bogolyubov coefficient related to the 
creation/annihilation operators and 
$p$ and $q$ are the out-going and in-going frequencies, 
respectively, that describe the modes used to expand the field subject to the accelerating boundary.

\subsection{Spectral Distribution} 

The spectral distribution \cite{Schwinger:1949ym} 
of the total radiation energy $E$ with 
respect to frequency $\omega$ and solid angle $\Omega$ is 
\be \frac{\diff{I(\omega)}}{\diff{\Omega}} \coloneqq \frac{\diff^2{E}}{\diff{\omega}\diff{\Omega}}\ ,\ee
see also \cite{Jackson:490457}. For the radiation of a moving point charge (in natural units with unit charge -- 
see e.g.\ Eq.~23.89 on page 911 of Zangwill \cite{Zangwill:1507229} in SI units or  
Eq.~14.67 on page 701 of Jackson \cite{Jackson:490457} in Gaussian units) 
this is given by the motion as 
\begin{equation}
	\frac{\diff I(\omega)}{\diff \Omega} = \frac{\omega^2}{16\pi^3}\,\left|\ \bd{\hat{n}} \times \int\displaylimits_{-\infty}^{\infty} d t\, \bd{\dot{r}}(t) e^{i\phi}\ \right|^2.
\label{density_zangwill}
\end{equation} 
Here $\omega$ is the frequency, $\bd{k} = \omega \bd{\hat{n}}$ the wave vector, $\d\Omega$ the 
solid angle, $\bd{r}$ the charge trajectory with 
velocity vector $\bd{\dot r}$, and 
$\phi =\omega t -\bd{k}\cdot\bd{r}(t)$. 
Defining $\bd{\hat{n}} \cdot \bd{\hat{r}} = \cos\theta$ and assuming straight line motion, 
we have 
\be 
\frac{\diff I(\omega)}{\diff \Omega} = \frac{\omega^2}{16\pi^3}\,\sin^2\theta\,\left|\ \int\displaylimits_{-\infty}^{\infty} d t\, \dot{r}(t) e^{i\phi}\ \right|^2 
\ .\label{eq:emit} 
\ee
Integrating this over solid angle 
$d\Omega=\sin\theta d\theta d\varphi$ and frequency $\omega$ will yield the total energy emitted. 

We can also interpret the trajectory as not 
that of a point charge but an accelerating 
mirror (boundary) and compare the horizon 
radiation from this dynamical Casimir effect. 
Thus we can test that the classical energy emitted 
agrees with the quantum result from the Bogolyubov 
creation/annihilation coefficients, and 
also, contrast  
the Larmor and Feynman powers. This further provides 
a way to derive the spectrum angular distribution 
for particle production from a moving mirror 
trajectory.

\subsection{Asymptotic Rest} \label{sec:endrest} 

To pursue an understanding of the spectrum 
angular dependence for the quantum analog, 
we consider moving mirror trajectories that  
deliver finite total 
energy and particle count (ensuring all integrals 
are convergent). 
Asymptotically inertial mirrors have finite total 
energy, while mirrors that also are asymptotically 
static 
(eventually coming to rest with zero velocity) have finite particle 
count, entropy, and have unitary evolution (seen geometrically since all light rays reflect off the mirror and none are lost). Therefore 
we consider only cases with asymptotic rest. 

The following list summarizes the only known trajectories possessing asymptotic rest with solved Bogolyubov coefficients. 

\begin{itemize} 
\item Walker-Davies \cite{Walker_1982}: but noninvertible $t(x)$. 
\item Arctx \cite{good2013time}: but nonfunctional particle count.
\item {\bf Self-Dual} \cite{Good:2017kjr}: time symmetric. 
\item {\bf betaK} \cite{Good:2018aer}: time antisymmetric. 
\item Schwarzschild-Planck \cite{Good:2019tnf,Good:2020fsw} (also see \cite{Moreno-Ruiz:2021qrf}): fully evaporating black hole with unitarity.
\end{itemize}

None of these have previously 
had published solutions for the beta 
Bogolyubov coefficients using both mirror sides to obtain the 3+1 D analog 
(and hence classical particle motion). 
In the next two sections 
we present solutions for the 
two boldface trajectories -- in particular as examples of time-symmetric vs antisymmetric motion, and the 
associated spectral distributions.

\section{Self-Dual Trajectory} \label{sec:selfdual} 

The self-dual mirror trajectory \cite{Good:2017kjr}
\be 
x(t)=\frac{-v}{\kp}\,\ln(\kp^2t^2+1)\ , \label{SDtraj}
\ee 
is even in time, and the self-dual nature means that 
the particle emission spectrum is equal on both sides of 
the mirror. The quantity $v$ is the maximum speed of 
the mirror, occurring at $\kp t=1$. The quantity $\kp$ 
sets the scale of the acceleration (and the surface 
gravity of the black hole analog in the 
accelerating boundary correspondence). 

The analog Larmor power radiated is 
\be 
P_L = \frac{2 \kappa ^2 v^2 \left(\kappa ^4 t^4-1\right)^2}{3 \pi  \left[\left(\kappa ^2 t^2+1\right)^2-4 \kappa ^2 t^2 v^2\right]^3}\ . \label{SDLarmor}
\ee 
As expected, no power is radiated by a 
stationary particle, $v=0$, and none at 
the moment of maximum velocity when the 
acceleration is zero (i.e.\ when $\kp t=1$, 
as well as at asymptotically early and late 
times). 

\begin{figure}[htbp]
\centering 
  \includegraphics[width=1.0\linewidth]{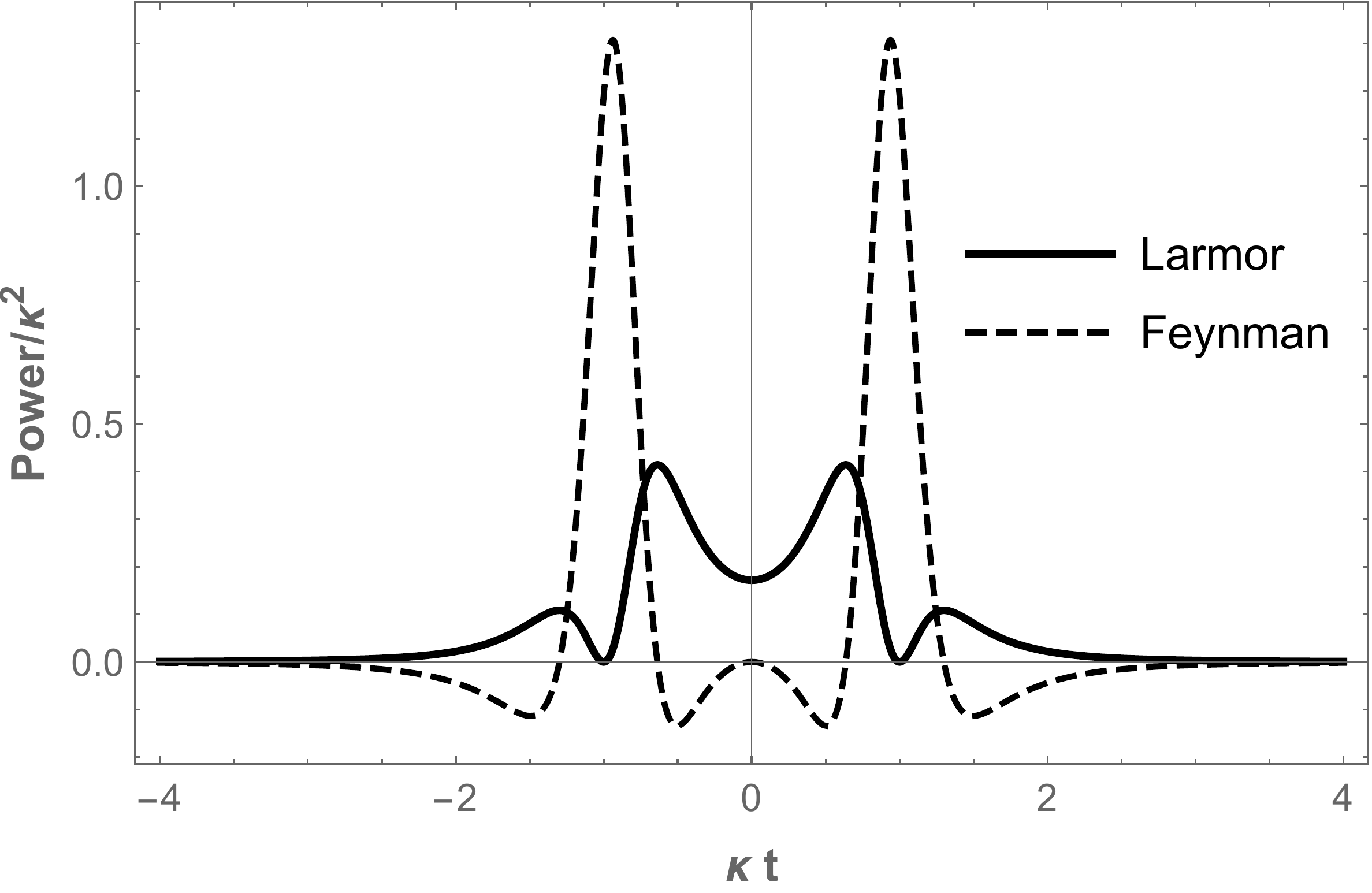}
 \caption{The Larmor and Feynman powers for the self-dual  trajectory are 
 plotted vs time, with $v=0.9$. 
A higher maximum velocity squeezes and heightens the peaks for both powers. 
 The Feynman power plotted is $P_F = - F \cdot v$ so that the total area under the curve is positive, $E = \int P_F \diff{t}$, see Eq.~(\ref{FL}).  
  Note the integrals under the curves are equal, giving the total energy radiated, 
Eq.~(\ref{SDenergy}). 
}  
\label{Fig_SDpower}
\end{figure}

The Feynman force can be similarly calculated 
analytically but the expression is long. 
Figure~\ref{Fig_SDpower} plots the Larmor and 
Feynman powers vs time. The Larmor power is of course 
always positive, while the Feynman power from the 
radiation reaction force can be both positive and 
negative. The Feynman power crosses zero at maxima 
of the Larmor power. Both types of power asymptotically vanish rapidly.

\begin{figure}[htbp]
\centering
  \includegraphics[width=1.0\linewidth]{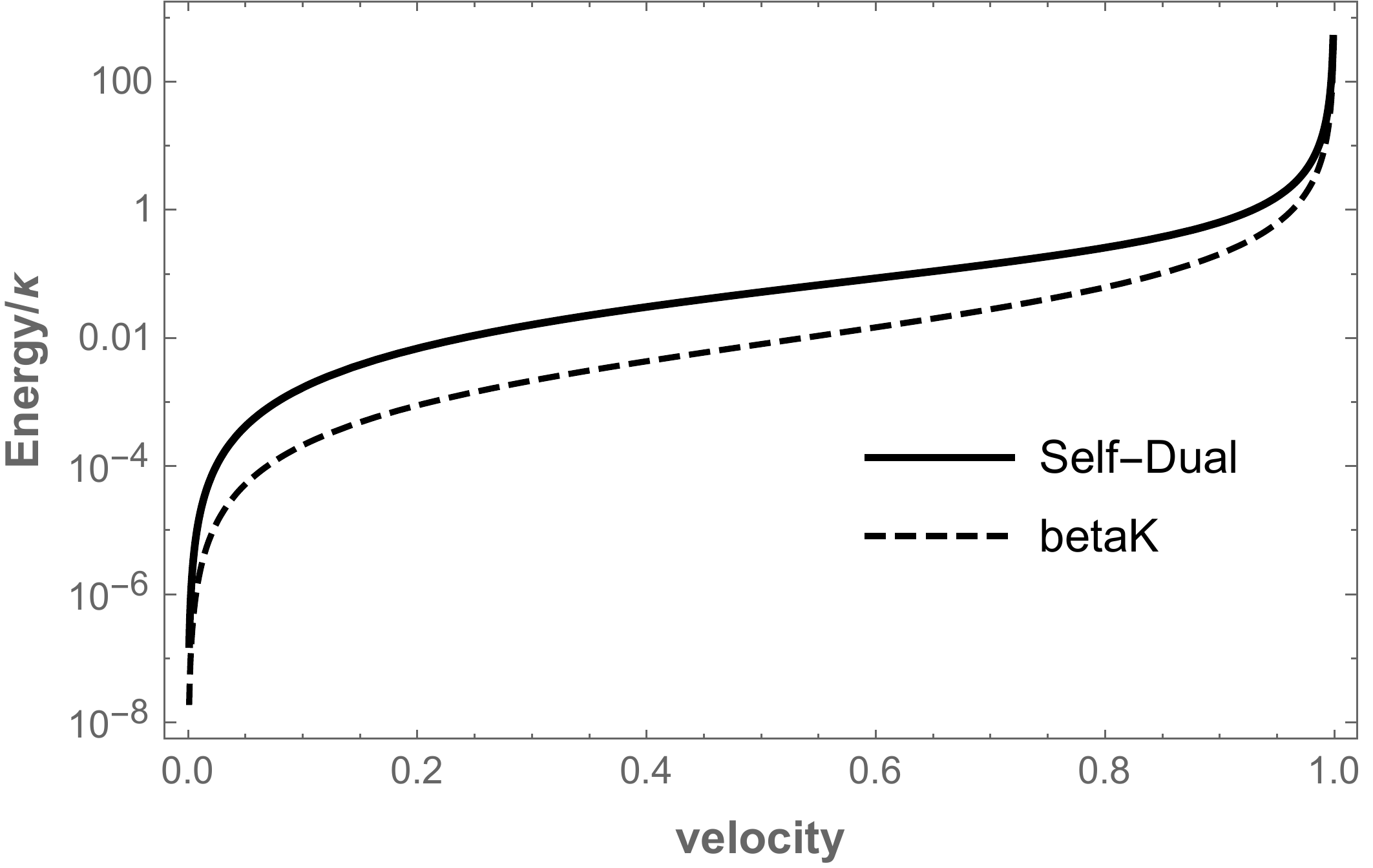}
\caption{The total energy 
as a function of maximum velocity parameter  is plotted for the self-dual trajectory 
(Eq.~\ref{SDenergy}) and the betaK trajectory 
(Eq.~\ref{energyBetaK}).}  
\label{fig:energy}
\end{figure}

Integrating over all time, Eq.~(\ref{FL}), the 
total energy emitted is 
\be 
E = \frac{\kappa}{24}\gamma v^2 \left(\gamma ^2+3\right)\ ,  \label{SDenergy}
\ee
where $\gamma=(1-v^2)^{-1/2}$ is the Lorentz factor. Figure~\ref{fig:energy} plots the 
total energy as a function of the maximum 
velocity. As the velocity approaches the speed 
of light, the Lorentz factor greatly increases 
the energy emitted.

\begin{figure}[htbp]
\centering 
\includegraphics[width=0.9\linewidth]{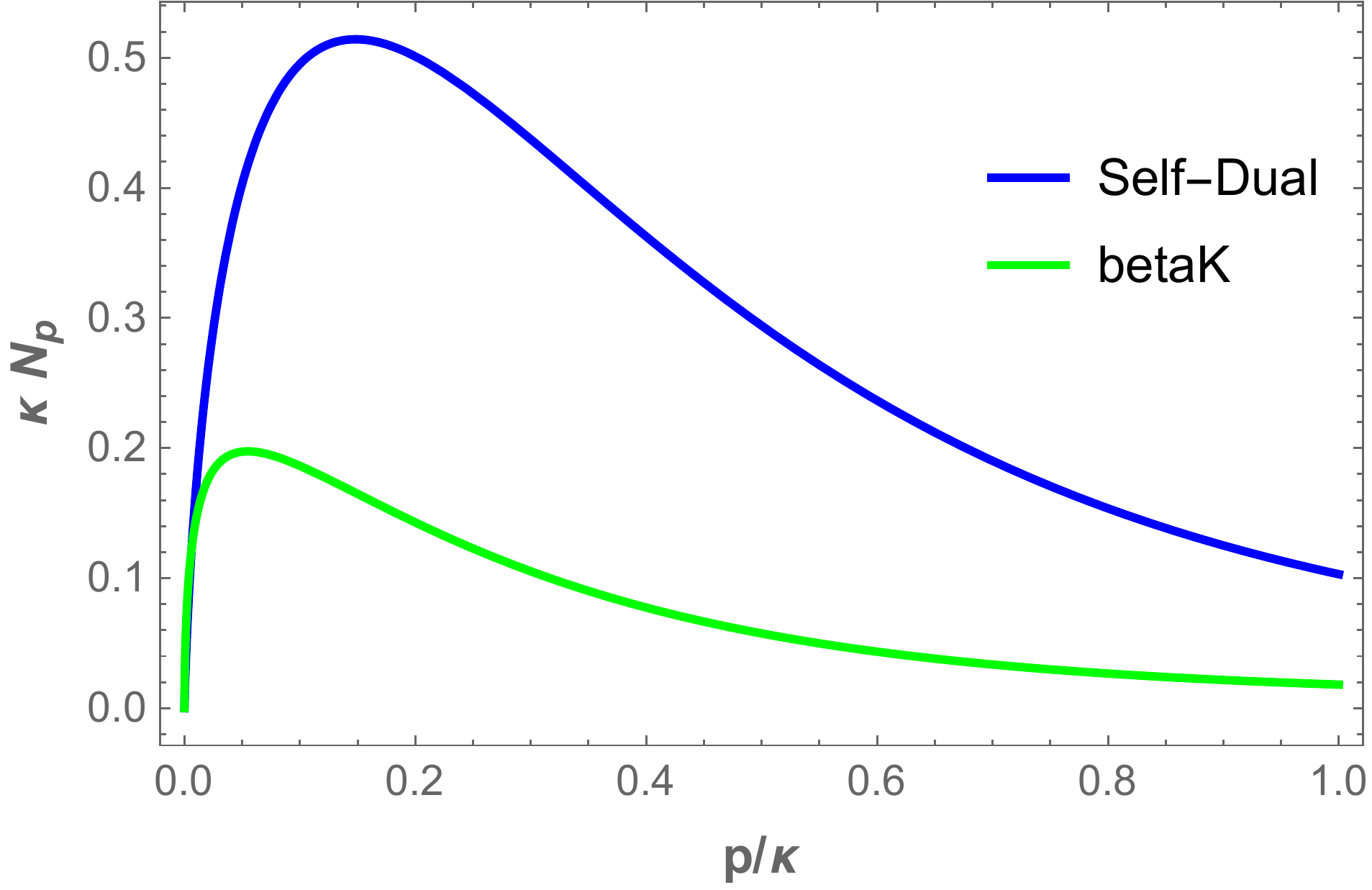}
 \caption{A plot of particle spectrum $N(p)$ from the mirrors. This is the particle count as a function of the outgoing mirror mode frequency, $p$. Here the maximum velocity of each mirror is $v=v_0 = 0.9$.  }  
\label{SDbK_Np_Fig}
\end{figure}

For the Bogolyubov spectrum as found from the double-sided moving mirror, the result 
(see e.g.\ \cite{good2013time} for the details 
of the steps) is 
\be 
|\beta_{pq}|^2 = \frac{16 v p q }{\pi ^2 \kappa ^2 \sigma \omega }\,\sinh\left(\frac{ \pi  v\sigma}{\kappa}\right)\,\left|K_{i v\frac{\sigma}{\kappa }+\frac{1}{2}}\left(\frac{\omega}{\kappa }\right)\right|^2 \ , \label{eq:bogSD}
\ee
where $\sigma = p - q$ and $\omega = p+q$. The particle spectrum $N_p=\int dq\, |\beta_{pq}|^2$ is non-thermal, and has finite particle production, as seen in Figure~\ref{SDbK_Np_Fig}.

For the spectral (angular) distribution, 
we use 
the self dual trajectory in Eq.~\eqref{eq:emit}, 
giving 
\be  
\frac{\diff{I}}{\diff{\Omega}} = \frac{v \omega^2}{\kappa^2 \pi^3} \frac{1-T^2}{2T} \sinh\left(\frac{\pi vT \omega}{\kappa}\right) \left|K_{\frac{1}{2}+\frac{i vT  \omega}{\kappa}}\left(\frac{\omega}{\kappa}\right)\right|^2\ , \label{SDdis}
\ee 
where $T\equiv\cos\theta$. 
Some details of the derivation are given in 
Appendix~\ref{sec:apxSDemit}. Note the 
similarity to the form of the beta Bogolyubov 
coefficients, but with added angular 
dependence (see the next subsection for further 
discussion). 

\begin{figure}[htbp]
\centering 
  \includegraphics[width=0.8\linewidth]{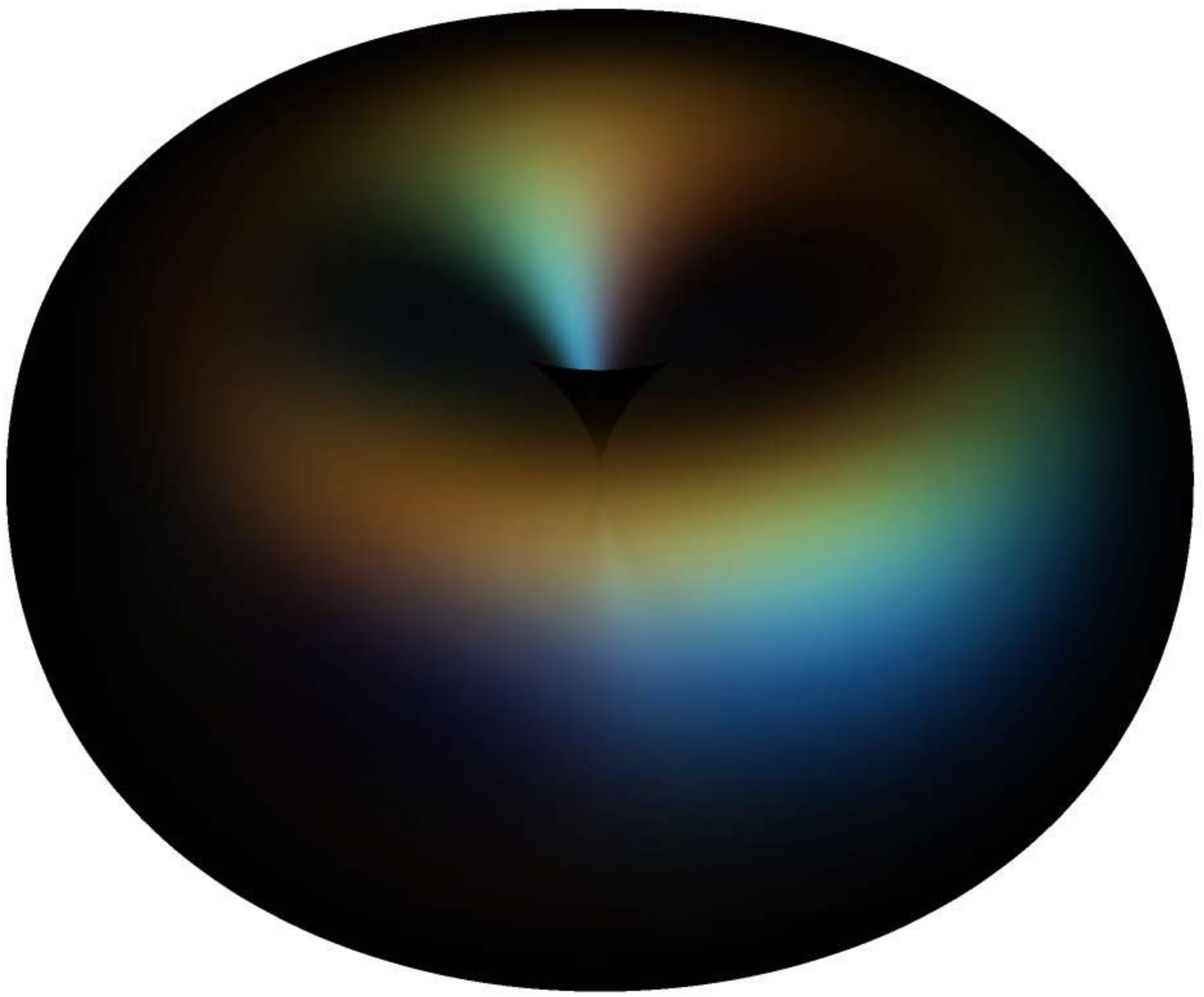}
 \caption{3D view of the radiated spectrum angular distribution $\diff I/\diff \Omega$ 
 from motion corresponding to the self dual trajectory. 
 Here we use unit charge, natural units, and $\omega = \kappa = 1$. The maximum 
 speed of the charge is $v= 0.95$. 
 Note the expected property of zero radiation directly in the forward direction. }  
\label{SD_distribution_Fig}
\end{figure}

\begin{figure}[htbp]
\centering 
  \includegraphics[width=0.5\linewidth]{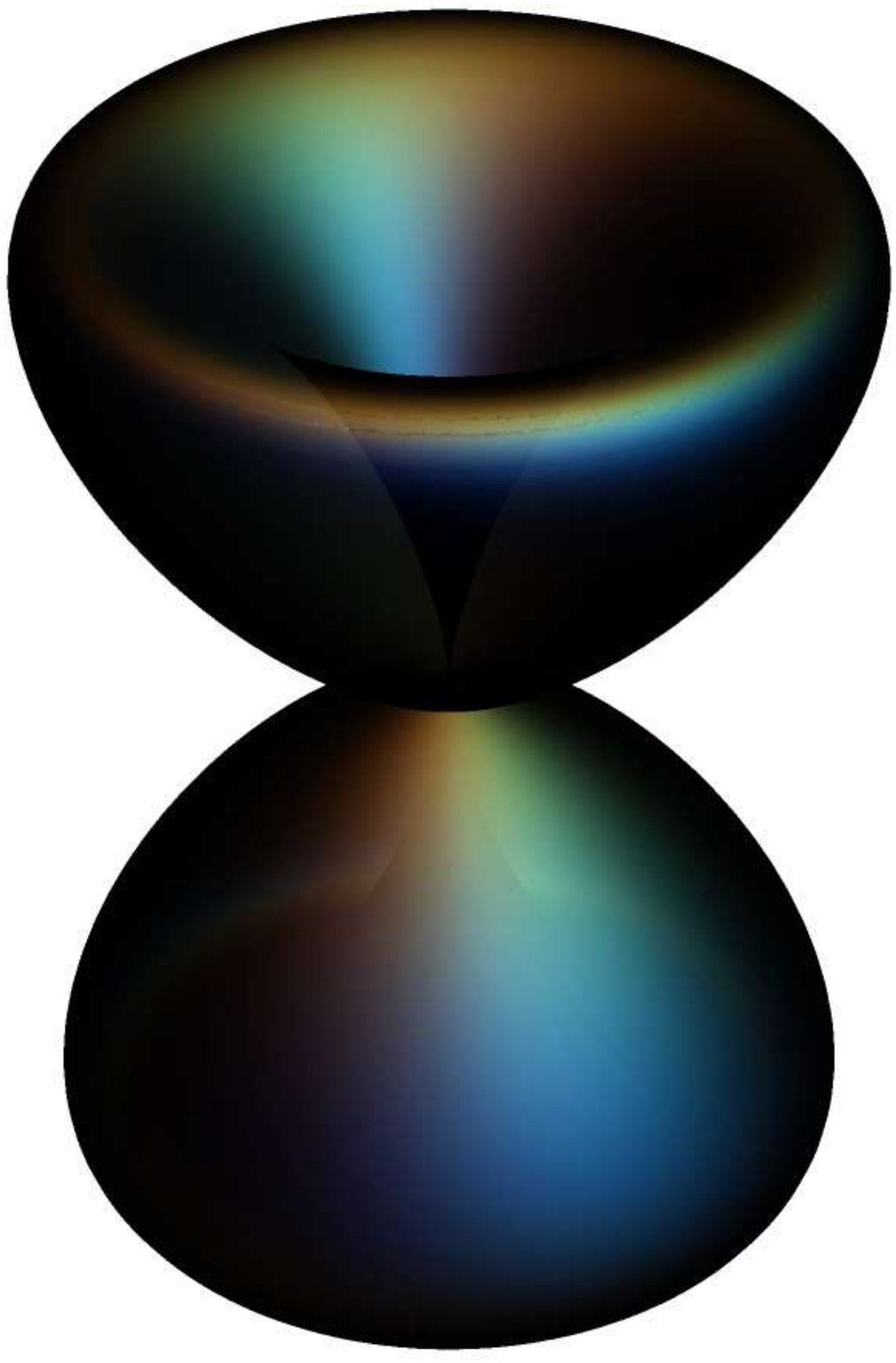}
 \caption{As Figure~\ref{SD_distribution_Fig} but 
 for $\omega =4$, $\kap=1$, showing the 
 high-frequency exponential suppression. 
}  
\label{SD_hourglass_Fig}
\end{figure}

\begin{figure}[htbp]
\centering 
  \includegraphics[width=0.9\linewidth]{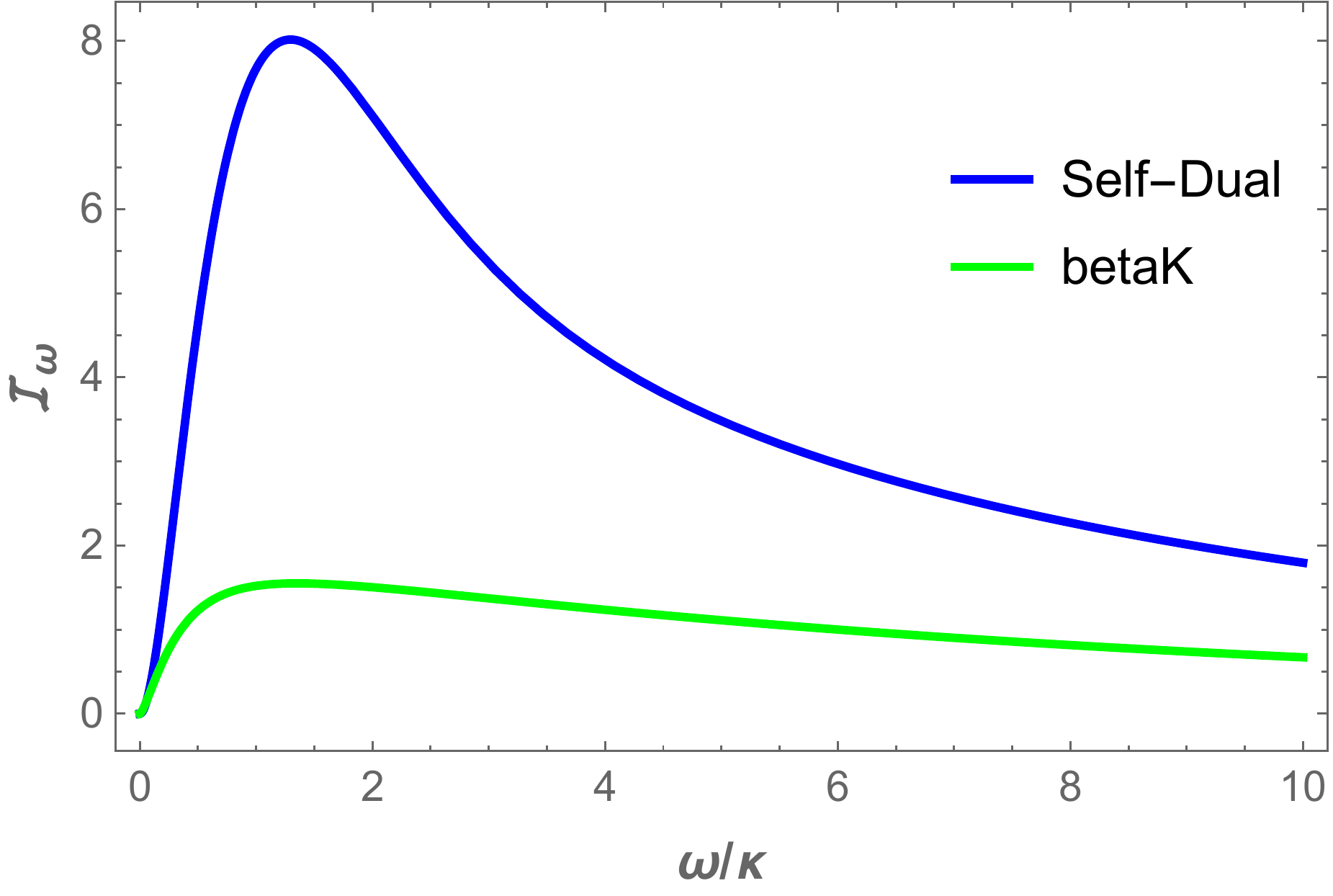}
 \caption{A plot of energy spectrum $I(\omega)$, which numerically integrates the spectral distributions for the self-dual, Eq.~(\ref{SDdis}), and betaK, Eq.~(\ref{bKdis}), 
 cases over solid angle $\Omega$. The vertical axis has been multiplied by $10^3$ for readability. Here the maximum velocity of each case is $v=v_0 = 0.9$.
 }
\label{SDbK_Iw_Fig}
\end{figure}

Figures~\ref{SD_distribution_Fig} and \ref{SD_hourglass_Fig} plot the spectral distribution in a 3D view. 
Notice there is no radiation in the forward or backward $T\to \pm 1$ ($\theta \to [0,\pi]$) directions.  This is expected of straight-line bremsstrahlung \cite{Griffiths:1492149}. The spectral distribution in the $T\to 0$ ($\theta \to \pi/2$) limit is:
\be \lim_{T\to 0}\,\frac{\diff{I}}{\diff{\Omega}} = \frac{v^2 \omega^2}{4 \pi  \kappa ^2}e^{-2\omega/\kappa}\,,\ee
which demonstrates a radiation allotment in directions perpendicular to the motion 
that is exponentially suppressed at high 
frequencies. 
The spectrum, $I(\omega)$, can be numerically found by integrating the spectral distribution, Eq.~(\ref{SDdis}), over solid angle.  
See Figure~\ref{SDbK_Iw_Fig} for an illustration.

The spectral distribution can be directly 
integrated over solid angle and 
frequency to obtain the total energy 
\bea 
E &=& \int_0^\infty \diff{\omega} \int_{-1}^{1}\diff{T} \int_0^{2\pi} \diff{\varphi}\; \frac{\diff{I}}{\diff{\Omega}}\\ 
&=& \frac{\kappa}{24}\,\gamma v^2 \left(\gamma ^2+3\right)\ . 
\eea 
This indeed agrees with Eq.~(\ref{SDenergy}).

\section{betaK Trajectory} \label{sec:betaK} 

The betaK trajectory \cite{Good:2018aer}
\be  
x(t)=\frac{-v_0}{\kp}\,\sinh^{-1}\kp t \label{eq:betaKtraj}\ ,  
\ee 
by contrast is odd in time, and 
gives more tractable solutions than the Walker-Davies 
or Arctx models. Furthermore it has an interesting 
relation to uniform acceleration in 3+1 D (though not 
in the 1+1 D mirror case)\footnote{We thank Ahmad 
Shariati for pointing this out.}. 
Its name arises because this trajectory has exactly solvable 
beta Bogolyubov coefficients involving a modified Bessel 
function $K$ in the moving mirror model, giving finite energy 
and finite particle production. 

This trajectory equation arises as well for a particle 
shot horizontally from the origin with an initial 
velocity $v_0$ (which is also the maximum velocity) encountering a constant 
vertical acceleration. Indeed, this is similar to the 
recently rediscovered ``Leonardo da Vinci's water pitcher'' that 
moves horizontally at constant speed $v$ spilling water 
in a uniform gravitational field \cite{10.1162/leon_a_02322} 
-- but here we consider relativistic speeds. 
The derivation appears in Appendix~\ref{sec:apxbetaK}. 

Note that in the relativistic case, despite 
no horizontal force the particles (water drops) do not have constant horizontal velocity:  
due to the coupling of horizontal and 
vertical motions through the Lorentz factor 
a horizontal acceleration is induced as made 
clear in Appendix~\ref{sec:apxbetaK}. 

\begin{figure}[htbp]
\centering 
  \includegraphics[width=1.0\linewidth]{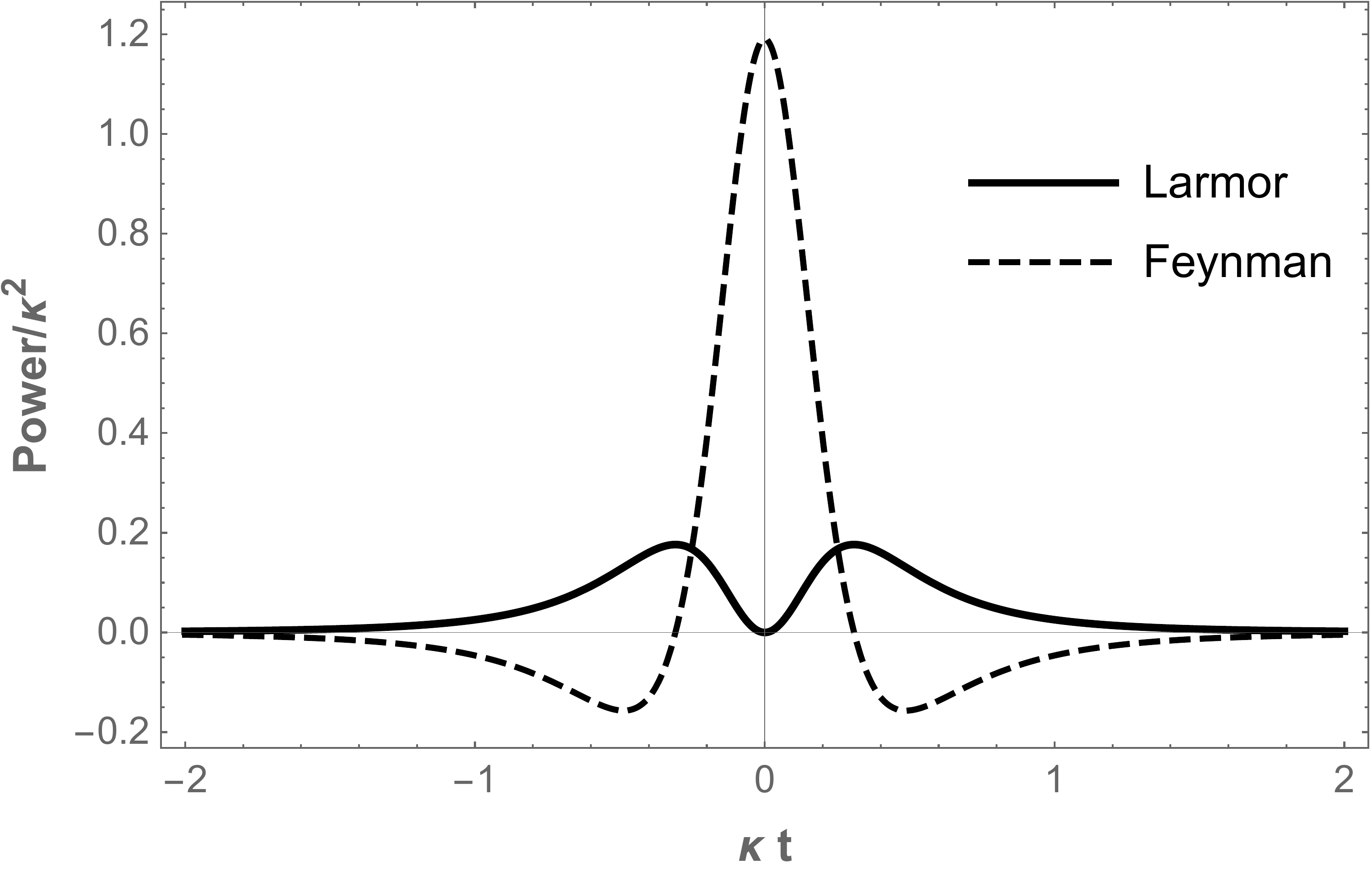}
 \caption{The Larmor and Feynman powers for the betaK trajectory are 
 plotted vs time, with $v_0 = 0.9$. Like the self-dual trajectory, a higher $v_0$ narrows and heightens the peaks for both powers. For illustration, the Feynman power plotted is $P_F = - F \cdot v$ so that the total area under the curve is positive.  The areas under the curves are equal, giving the total energy radiated, 
 Eq.~(\ref{energyBetaK}).}  
\label{Fig_betaKpower}
\end{figure}

The Larmor power radiated by a charge with the 
betaK trajectory is 
\be 
P_L = \frac{\alpha^2}{6\pi}=\frac{\kappa ^2}{6\pi}\gamma^6 \left(v_0^2-V^2\right)\frac{V^4}{v_0^4}\ , \label{BKLarmor}
\ee 
where the velocity is 
\be 
V(t)\equiv\dot x(t)=\frac{-v_0}{\sqrt{\kp^2 t^2+1}}\ . 
\ee 
The speed $|V|\le |v_0|$ so the power always remains nonnegative. For this time 
antisymmetric trajectory, the power has only 
one maximum on each side of $t=0$ and no 
zeros for finite $t\ne0$. 
The Feynman power is 
\be 
P_F=\frac{\alpha^2}{6\pi}\,\left[2-\frac{V^2(1-v_0^2)}{v_0^2-V^2}\right]\ . 
\label{BKFeynman} 
\ee 
The total energy, using Eq.~(\ref{FL}), is
\be 
E = \frac{\kappa}{48}\gamma_0^3 v_0^2\ .\label{energyBetaK} 
\ee 
See Figure \ref{fig:energy} for the energy and Figure~\ref{Fig_betaKpower} for the Larmor and Feynman powers.

The Bogolyubov spectrum as found from the double-sided moving mirror is
\be 
|\beta_{pq}|^2 = \frac{8 v_0^2 p q }{\pi ^2 \kappa ^2 \omega^2 }\cosh\left( \pi  v_0 \frac{ \sigma }{\kappa }\right)\left|K_{i v_0\frac{\sigma}{\kappa }}\left(\frac{\omega}{\kappa }\right)\right|^2\ , 
\ee
where $\sigma = p-q$ and $\omega = p+q$. 
This spectrum is not thermal. 
Note the similarities, but also subtle differences 
with the self-dual case, Eq.~\eqref{eq:bogSD}. The energy is confirmed by associating a quantum $\hbar p$ (where $p$ is the outgoing frequency mode) and integrating using Eq.~(\ref{energyfrombeta}), which yields Eq.~(\ref{energyBetaK}). 
The particle spectrum $N_p=\int dq\, |\beta_{pq}|^2$ is shown in 
Figure~\ref{SDbK_Np_Fig}. 

Using the betaK trajectory within classical electrodynamics \cite{Jackson:490457}, we find the spectral distribution, 
\be \frac{\diff I}{\diff \Omega} = \frac{v_0^2 \omega^2}{4\kappa^2\pi^3}(1-T^2) \cosh\left(\pi v_0 T\frac{\omega}{\kappa}\right)\left|K_{i v_0 T \frac{\omega}{\kappa}}\left(\frac{\omega}{\kappa}\right)\right|^2.\label{bKdis}\ee
where $T=\cos\theta$. 
Again a relation between the classical 
spectral distribution and quantum beta Bogolyubov 
coefficient is apparent; we address this 
in Section~\ref{sec:disc}. 

\begin{figure}[htbp]
\centering 
  \includegraphics[width=0.8\linewidth]{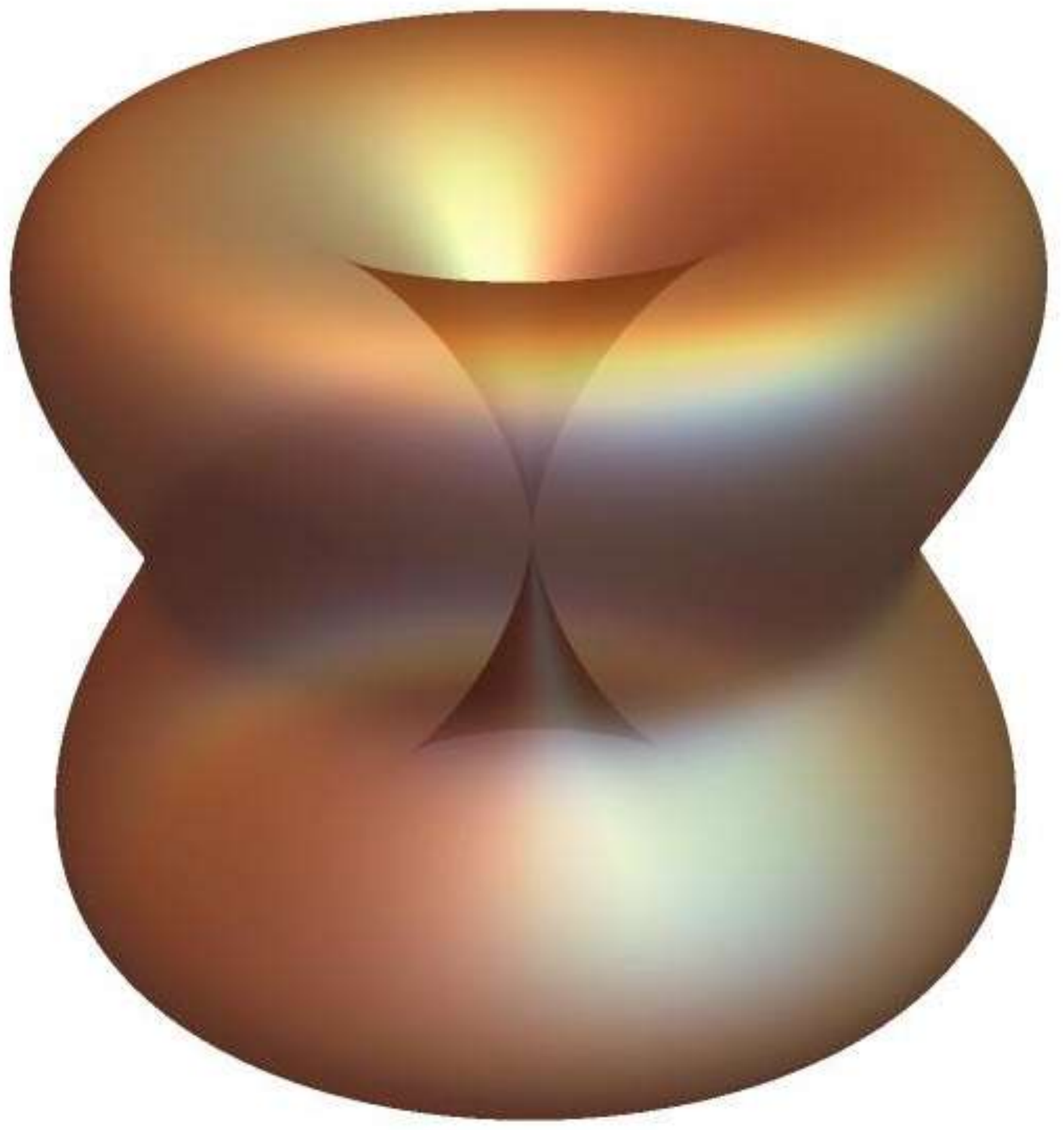}
 \caption{3D view of the radiated spectrum  
angular distribution $\diff I/\diff \Omega$ 
from motion corresponding to the betaK 
trajectory. 
Here we use unit charge, natural units, and $\omega = \kappa = 1$. The maximum 
 speed of the charge is $v_0= 0.95$. 
 }  
\label{bK_distribution_Fig}
\end{figure}

The energy spectrum $I(\omega)$ is shown in 
Figure~\ref{SDbK_Iw_Fig}. 
Integration of Eq.~\eqref{bKdis} over $\diff{\omega}\diff{\Omega}$ agrees with the 
total energy of Eq.~(\ref{energyBetaK}). 
Like the self-dual case, there is no radiation in the forward or backward $T\to \pm 1$ ($\theta \to [0,\pi]$) directions, as expected. See 
Figure~\ref{bK_distribution_Fig} for a 3D view of the spectral distribution. The spectral distribution in 
the $T\to 0$ ($\theta \to \pi/2$) limit is:
\bea 
\lim_{T\to 0}\,\frac{\diff{I}}{\diff{\Omega}} &=& \frac{v_0^2 \omega^2 }{4 \pi ^3 \kappa^2}\, \left[K_0\left(\frac{\omega}{\kappa}\right)\right]^2\\
&\approx& \frac{v_0^2 \omega  }{8 \pi ^2 \kappa }\,e^{-\frac{2 \omega }{\kappa }}\ ,\eea
again showing the high-frequency exponential 
suppression, wherein the second line we have expanded around large $\omega/\kp$.  

The betaK trajectory is well-motivated, physically intuitive, and potentially realizable in the laboratory 
as it is straightforwardly the horizontal component of an electron's motion subject to an initial horizontal velocity and constant vertical force.  In the following section, we use betaK's analytic tractability to help confirm the duality between the classical point charge and the quantum moving mirror.

\section{Classical-Quantum Correspondence} \label{sec:disc} 

\begin{figure}[htbp]
\centering 
  \includegraphics[width=0.9\linewidth]{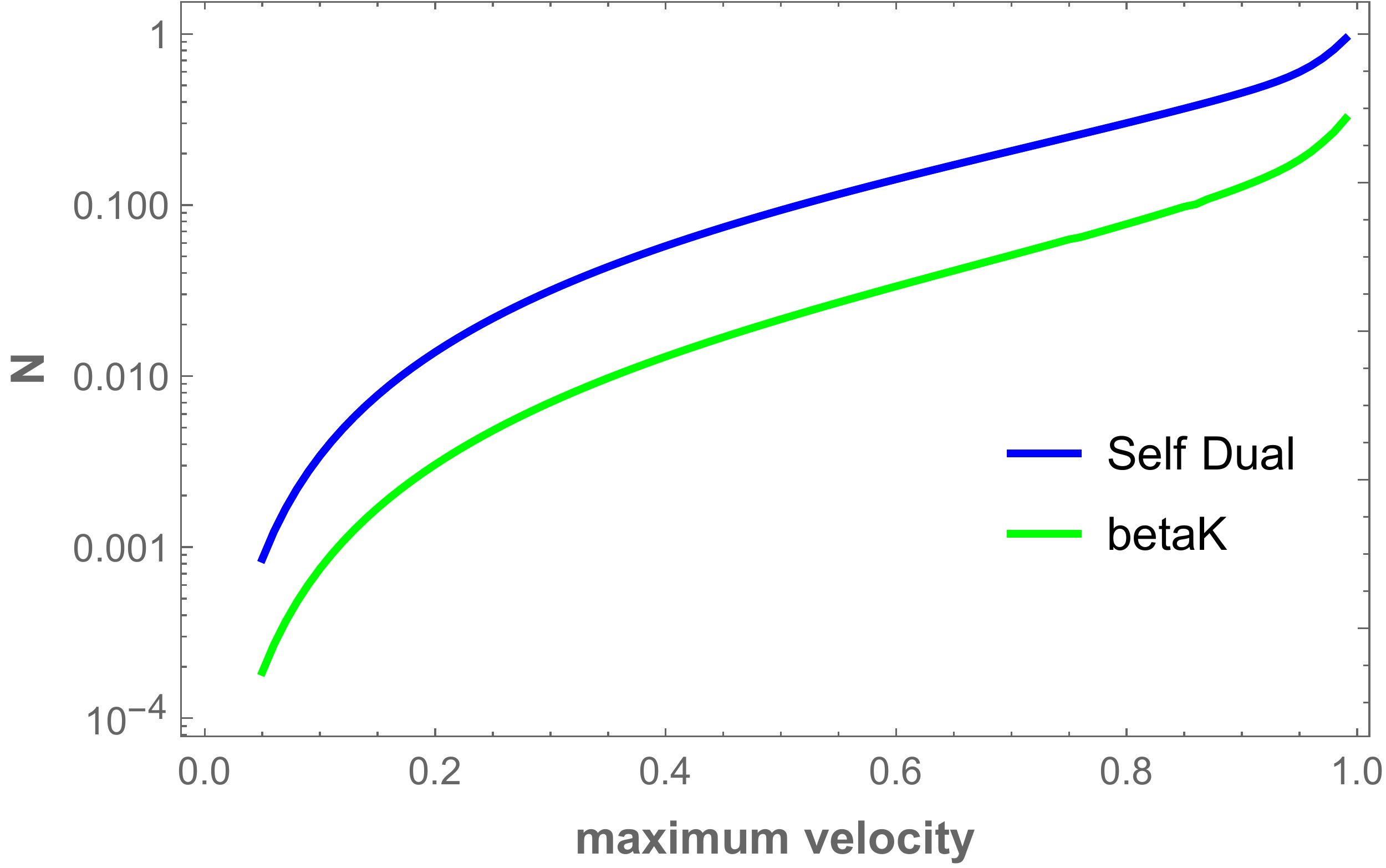}
 \caption{A plot of total finite particle count of the radiation particles created by the two mirrors, using Eq.~(\ref{particles}), for maximum velocity 
 ranging from 0.05 to 0.99. } 
\label{SDbK_particles_Fig}
\end{figure}

We have seen that at the level of total energy there 
is agreement between the charge radiation approach and 
the moving mirror Bogolyubov coefficient approach, 
\be 
E=\int_0^\infty \diff{\omega} \int_{-1}^{1}\diff{T} \int_0^{2\pi} \diff{\varphi}\; \frac{\diff{I}}{\diff{\Omega}}
\Leftrightarrow \int_0^\infty \int_0^\infty p\, |\beta_{pq}|^2 \diff{p}\diff{q}\ . 
\ee 
We can further see that the agreement extends to 
the particle count, 
\be 
N = \int \frac{1}{\omega} \frac{\diff{I}}{\diff{\Omega}} \diff{\Omega}\diff{\omega} 
\Leftrightarrow \frac{1}{2}\int\int |\beta_{pq}|^2 \diff{p}\diff{q} \ . \label{particles}
\ee 
The factor $1/\omega$ converts particle energy 
to particle number, and the factor 1/2 arises because 
while both sides of the mirror are employed in the 
correspondence, an observer could only see one side 
See Figure~\ref{SDbK_particles_Fig} for an illustration of particle count.

As mentioned in Section~\ref{sec:selfdual} and 
Section~\ref{sec:betaK} the connection persists 
at the level directly between the spectral distribution and 
the beta Bogolyubov coefficient, i.e.\ the integrands. 
The steps to obtain the exact relation are as follows. 
First, the Jacobian going from $\{p,q\}$ coordinates 
to $\{\omega,T\}$ coordinates 
is $\omega/2$. Recall that $\diff{\Omega}=\sin\theta \diff{\theta}\diff{\varphi}$ 
and that $\diff{T} \equiv \diff{(\cos\theta)}=\sin\theta \diff{\theta}$ and the 
$\diff{\varphi}$ integral simply contributes $2\pi$. 
Finally, the parity is reversed on opposite sides 
of the mirror so that one side is related to the 
other by $T\leftrightarrow -T$, so we write 
\be   
\int_{-1}^{+1}dT\,\frac{dI}{d\Omega}=\frac{1}{2}\left[\int_{-1}^{+1}dT\,\frac{dI(T)}{d\Omega}+\int_{-1}^{+1}dT\,\frac{dI(-T)}{d\Omega}\right]\ . 
\ee 

Putting all the elements together delivers the 
correspondence 
\be 
|\beta_{pq}|^2 \ \leftarrow\ \frac{4\pi}{\omega^2}\left[\frac{\diff{I}}{\diff{\Omega}}(\omega,\cos\theta) + \frac{\diff{I}}{\diff{\Omega}}(\omega,-\cos\theta)\right]\ . 
\ee 
This can be verified directly for the solutions given for the two trajectories. Note that the correspondence formally goes in 
only one direction, from charge radiation to moving 
mirror, as the beta Bogolyubov coefficient has no 
angular information on the in-going and out-going 
modes. Only once we introduce an angle $\theta$ such 
that $p=\omega(1+\cos\theta)/2$ and 
$q=\omega(1-\cos\theta)/2$, hence $p+q=\omega$ and 
$\sigma\equiv p-q=\omega\cos\theta$, 
can we go the other way. 

Such a classical-quantum correspondence is very 
useful, but we emphasize that it does not capture 
all quantum effects. While the particle production 
can be computed classically, this neglects 
quantum effects when the radiation (photon) energy 
becomes comparable to the particle (electron) energy, 
e.g.\ the radiation wavelength is smaller than the 
charge de Broglie wavelength.

\section{Conclusions} \label{sec:concl} 

We have solved for the accelerating point charge radiation 
-- its energy, particle count, and spectral angular 
distribution -- of two trajectories that asymptotically 
come to complete stop, compatible with finite total 
particle emission. As Feynman \cite{Feynman:1996kb} has emphasized, 
\\\\
\textit{Larmor's power is only valid for cyclic motions, or at least motions which do not grow forever in time.} 
\\\\
The betaK and Self-Dual trajectories fulfill that condition, 
and these two solutions inspired by the accelerating boundary 
(moving mirror) analog are the only known rectilinear solutions 
with exactly soluble spectra, finite energy, and finite particle count. This allows comparison of classical and quantum systems 
directly. 

The main results presented include: 

\begin{itemize}
    \item We have found the time dependence of radiative solutions. One utility of an exact solution for moving point charge radiation is that in QED, time-dependent computations are notoriously difficult. Here the dynamics are explicit in the applicable Larmor and Feynman powers. 
\item We have demonstrated consistency between the total energy 
derived in terms of the Larmor power, the Feynman power, and 
the quantum Bogolyubov coefficients. 
    \item We have derived the spectral distributions of these two accelerating, but asymptotically static, motions analytically, and further shown consistency with the total energy emission and total particle count. In addition to 3D plots of the radiation 
    angular distribution we discussed the angular limits (e.g.\ the 
    forward and transverse emission) and high frequency limits. 
    \item We have laid out explicitly a quantum-classical correspondence to the moving mirror model, mapping between 
    the classical spectral distribution and the quantum 
Bogolyubov coefficients.
\end{itemize}

The demonstrated consistency and explicit correspondence 
enhances the utility of the moving mirror model by 
showing its role as a point charge analog. Thus the 
accelerated boundary correspondence of the moving mirror 
to black hole radiation may potentially point to a 
connection to accelerating charge radiation via a 
Hawking-Feynman-Larmor correspondence. 

This is an exciting prospect for future directions. 
It may be tractable to link directly these electron trajectories to curved spacetime counterparts, revealing 
spacetime metrics that radiate with similar nonthermal spectra 
(or reveal charge motions that could show a period of thermal 
emission). 
Further, given that a connection for beta decay to a moving 
mirror analog has been made 
\cite{Good:2022eub,Lynch:2022rqx,Good:2022xin}, other 
well-known QED scattering processes might correspond at lowest 
order to one of the solutions given. 
Asymptotic rest, with its finite particles and unitarity, 
could be a powerful tool, and it would be interesting to 
develop further solutions, such as 
the Schwarzschild-Planck radiation \cite{Good:2019tnf,Good:2020fsw,Moreno-Ruiz:2021qrf} to 
compare accelerating electron and black hole radiation in 
the thermal limit.

\acknowledgments 

Funding comes in part from the FY2021-SGP-1-STMM Faculty Development Competitive Research Grant No.\ 021220FD3951 at Nazarbayev University. This work is supported in part by the Energetic Cosmos Laboratory, and in part by the U.S.\ Department of Energy, Office of Science, Office of High Energy Physics, under contract no.\ DE-AC02-05CH11231.

\appendix

\section{Spectral Distribution Calculation} \label{sec:apxSDemit} 

To show how one can go from the formula for the spectral distribution, 
Eq.~\eqref{eq:emit}, to the modified Bessel function 
result we illustrate the steps for the self dual case. 
The integral has the form 
\be 
A\equiv\int_{-\infty}^{+\infty} dt\,\dot x\,e^{i\omega(t-x\cos\theta)}\ . 
\ee 
Substituting in the self dual expressions for $x(t)$ 
from Eq.~\eqref{SDtraj}, and $\dot x$, and writing $T\equiv\cos\theta$ 
we have 
\bea 
A&=&\int_{-\infty}^{+\infty} dt\,\frac{-2v\kp t}{\kp^2 t^2+1}\,
e^{i\omega[t+(vT/\kp)\ln(\kp^2 t^2+1)]}\\
&=&\frac{-2v}{\kp}\int_{-\infty}^{+\infty} ds\,s(s^2+1)^{-1+i\omega vT/\kp}\,
e^{i\omega t}\\ 
&=&\frac{-4iv}{\kp}\int_{0}^{\infty} ds\,s(s^2+1)^{-1+i\omega vT/\kp}\,
\sin\frac{\omega s}{\kp}\ . 
\eea
In the second line we have taken the exponential of the log term,
and defined $s=\kp t$, while in the third line we have used that we
must take the odd part of the remaining exponential to give an even
integrand over the symmetric range of integration.

This integral can be evaluated through Gradshteyn \& Ryzhik 3.771.5 \cite{gradshteyn1994}, 
resulting in 
\bea
A&=&\frac{4v}{\kp\sqrt{\pi}}\,\left(\frac{\omega}{2\kp}\right)^{1/2-i\omega vT/\kp}\,\sinh(\pi\omega vT/\kp)\,\Gamma\left(\frac{i\omega vT}{\kp}\right)\notag\\ 
&\quad&\times K_{1/2+i\omega vT/\kp}\left(\frac{\omega}{\kp}\right)\ .
\eea 
The modulus squared, using that $|\Gamma(ix)|^2=\pi/(x\sinh x)$, 
is
\be
|A|^2=\frac{8v}{\kp^2 T}\,\sinh(\pi\omega vT/\kp)\,
\left|K_{1/2+i\omega vT/\kp}\left(\frac{\omega}{\kp}\right)\right|^2\ .
\ee 

For the betaK case we proceed similarly, noting that 
since $\dot x$ is even in time in that case we must 
take the even part of the exponential (i.e.\ cosine).

\section{Leonardo's Pitcher: From Electron to betaK} \label{sec:apxbetaK} 

The motion of a relativistic particle with unit mass 
subject to an 
external force comes from the action\footnote{This 
is a first prototypical system of a relativistic Lagrangian 
(see e.g.\ page 323 of \cite{goldstein:mechanics}).} 
\be 
S =  -\int dt\; \left(\sqrt{1-v^2} + F x\right)\ . 
\ee 
For a force dependent only on position the 
equations of motion are simply 
\bea 
\alpha&=&\frac{d}{dt}\frac{v}{\sqrt{1-v^2}}\equiv\frac{d(\gamma v)}{dt} \label{eom1} \\ 
&=&(0,\alpha_y,0)\ , 
\eea 
where the last line holds for purely vertical force, 
and we will take $\alpha_y=\,$const 
(e.g.\ gravity in Leonardo's water pitcher experiment). 
Finally, we take the initial velocity to be purely 
horizontal, $v=(v_0,0,0)$. 

The results are simple -- nonuniform motion in the 
horizontal direction due to the relativistic boost 
factor $\gamma$, and hyperbolic motion under constant 
acceleration in the vertical 
direction -- but worth quickly going through to reveal 
the form of nonuniformity. 

The $z$ direction is trivial: as there is no 
initial velocity, nor subsequent acceleration, in this 
direction then Eq.~\eqref{eom1} guarantees that $z(t)=z(0)$ 
and we can ignore this dimension. In the $x$ (horizontal) 
direction, Eq.~\eqref{eom1} gives 
\be 
\gamma(t) v_x(t) = \gamma_0 v_0\ , \label{coupling1} 
\ee 
and the key point is that while nonrelativistically 
one would simply have $v_x(t)=v_0$, i.e.\ uniform 
motion, the Lorentz factor $\gamma$ couples in the 
$y$ motion (recall $\gamma=1/\sqrt{1-v_x^2-v_y^2}$), 
which is accelerated. This results in 
nonuniform motion horizontally. 

We can relate $v_x$ and $v_y$, and solve for both 
motions by squaring Eq.~\eqref{coupling1} to get 
\be 
v_x^2 = (1-v_y^2)v_0^2\ . \label{coupling2} 
\ee 
This immediately tells us that $v_x$ has its maximum 
value at the initial time, so $v_y(t)<v_0=v_y(0)$. 
That is, the vertical acceleration effectively causes 
a horizontal deceleration! 

In the $y$ (vertical) direction, the equation of 
motion gives $\gamma v_y = \alpha_y t$ so 
\be 
v_y = \frac{\kappa t}{\sqrt{1+(\kappa t)^{2}}}\ . \label{vx} 
\ee
At late times this approaches the speed of light. 
To presage the betaK mirror analogy we have written 
$\kappa\equiv\alpha_y/\gamma_0$. Finally, with 
Eq.~\eqref{coupling2} we obtain the horizontal velocity
\be 
v_x = \frac{v_0}{\sqrt{1+(\kappa t)^2}}\ , \label{vy} 
\ee 
which indeed decelerates from its initial value to zero. 
Again presaging the mirror analog, we will end up with 
an asymptotically static mirror defined by the 1D horizontal motion. 

Integrating the velocities gives the trajectories, with 
\be 
y(t)=\kappa^{-1}\sqrt{1+\kappa^2 t^2} -\kappa^{-1}\ , \label{hyperbola} 
\ee
revealing hyperbolic motion in the vertical direction. 
In the horizontal direction, 
\be 
x(t) = \frac{v_0}{\kappa}\, \sinh^{-1}\kappa t\ , \label{betaK1} 
\ee
exactly (after a trivial sign flip on initial velocity) 
the betaK trajectory, Eq.~(\ref{eq:betaKtraj}).





\bibliography{main} 
\end{document}